\documentclass[10pt,conference, anonymous]{IEEEtran}
\IEEEoverridecommandlockouts
\usepackage[noend]{algpseudocode}
\usepackage{amsmath,amssymb,amsfonts}
\usepackage[labelformat=simple]{subcaption}

\usepackage[hidelinks]{hyperref}
\usepackage[utf8]{inputenc}
\usepackage{dblfloatfix}
\usepackage{tcolorbox}
\usepackage{algorithm}

\usepackage{graphicx}
\usepackage{booktabs}
\usepackage{textcomp}
\usepackage{graphicx}
\usepackage{cleveref}
\usepackage{multirow}
\usepackage{textcomp}
\usepackage{balance}
\usepackage{comment}
\usepackage{siunitx}
\usepackage{diagbox}
\usepackage{xcolor}
\usepackage{pifont}
\usepackage{bm}

\newcommand{\cmark}{\checkmark}

\usepackage[
backend=biber
,style=ieee
,minbibnames=1
,maxbibnames=99
,maxcitenames=1
,mincitenames=1
,eprint=false
,doi=false
,isbn=false
,url=false
]{biblatex}
\bibliography{ref} 
\AtBeginBibliography{\footnotesize}

\begin{document}

\title{Using Source Code Metrics for Predicting Metamorphic Relations at Method Level}

\author{\IEEEauthorblockN{
	Alejandra Duque-Torres\IEEEauthorrefmark{2},
	Dietmar Pfahl\IEEEauthorrefmark{2}, 
    Claus Klammer\IEEEauthorrefmark{3}, and
    Stefan Fischer\IEEEauthorrefmark{3}}
			
	\IEEEauthorblockA{%
	\IEEEauthorrefmark{2}\textit{Institute of Computer Science  }, \textit{University of Tartu}, Tartu, Estonia \\
	E-mail: \{duquet, dietmar.pfahl\}@ut.ee}	
	\IEEEauthorblockA{
	\IEEEauthorrefmark{3}\textit{Software Competence Center Hagenberg (SCCH) GmbH}, Hagenberg, Austria \\
	E-mail:  \{claus.klammer, stefan.fischer\}@scch.at}
}

\maketitle

\begin{abstract}
Metamorphic testing (TM) examines the relations between inputs and outputs of test runs. These relations are known as metamorphic relations (MR). Currently, MRs are handpicked and require in-depth knowledge of the System Under Test (SUT), as well as its problem domain. As a result, the identification and selection of high-quality MRs is a challenge. \citeauthor{PMR1} suggested the Predicting Metamorphic Relations (PMR) approach for automatic prediction of applicable MRs picked from a predefined list. PMR is based on a Support Vector Machine (SVM) model using features derived from the Control Flow Graphs (CFGs) of 100 Java methods. The original study of \citeauthor{PMR1} showed encouraging results, but developing classification models from CFG-related features is costly. In this paper, we aim at developing a PMR approach that is less costly without losing performance. We complement the original PMR approach by considering other than CFG-related features. We define 21 features that can be directly extracted from source code and build several classifiers, including SVM models. Our results indicate that using the original CFG-based method-level features, in particular for a SVM with random walk kernel (RWK), achieve better predictions in terms of AUC-ROC for most of the candidate MRs than our models. However, for one of the candidate MRs, using source code features achieved the best AUC-ROC result (greater than 0.8).

\end{abstract}


\section{Introduction}
\label{sec:intro}

\textit{Metamorphic Testing} (MT) is a software testing technique that attempts to alleviate the test oracle problem \cite{chen2020metamorphic}. A test oracle is a mechanism for determining whether or not the outcomes of a program are correct \cite{QuASoQ,6963470}. The oracle problem arises when the system under test (SUT) lacks an oracle or when developing one to verify the computed results is practically impossible \cite{6963470}. Instead of verifying the individual outputs of the SUT, as traditional software testing techniques do, MT examines the relations between the inputs and outputs of test runs. These relations are known as Metamorphic Relations (MRs). MRs define how outputs should vary in response to a defined change of inputs when executing the SUT \cite{6613484,8493260}. If a particular MR is violated for at least one test input (and its change), there is a high probability that the SUT has a fault. On the other hand, if a particular MR is not violated it does not guarantee that the SUT is free of flaws. Thus, the effectiveness of MT is greatly dependent on the appropriateness of the MRs used \cite{6613484}.

Identifying and selecting appropriate MRs is not a trivial task. It requires a deep understanding of the SUT and its domain. Identifying and selecting high-quality MRs has been noted as a significant challenge. Several approaches to determine how to choose ``good" MRs have been proposed. For instance, \citeauthor{6319226}~\cite{6319226} introduced the Composition of MRs (CMRs) technique for constructing new MRs by mixing multiple existing ones. \citeauthor{10.1145/2642937.2642994}~\cite{10.1145/2642937.2642994} proposed a method in which an algorithm searches for MRs expressed as linear or quadratic equations. \citeauthor{CHEN2016177}~\cite{CHEN2016177} developed METRIC, a specification-based technique and related tool for identifying MRs based on the category-choice framework. They also expanded METRIC into METRIC+ by integrating the information acquired from the output domain \cite{8807231}. Among those approaches, Predicting Metamorphic Relations (PMR),  introduced by \citeauthor{PMR1}~\cite{PMR1, PMR3}, uses machine learning (ML) techniques to automatically detect likely MR of program methods using features extracted from a method’s control-flow graph (CFG). 

The idea behind the PMR approach is to build a model that predicts whether a method in a newly developed SUT can be tested using a specific MR. The key part of PMR is \textit{feature design}. In the original PMR work \cite{PMR1}, \citeauthor{PMR1} used the CFG's path- and node-based features of 48 Java methods and three predefined MRs to train support vector machines (SVM) and decision trees models. The authors show encouraging results when using node- and path-based features and SVM. Then, \citeauthor{PMR3}~\cite{PMR3} extends their initial work by examining 100 Java methods and a set of six predefined MRs. As in the initial study, \citeauthor{PMR3} use SVMs, and features extracted from the methods' CFGs. However, instead of the node- and path-based features, the authors used measures of similarity between graphs. In particular, random walk kernel (RWK) and graphlet kernel (GK). \citeauthor{PMR3} concluded that SVM models trained with RWK-based features performed better than those trained with GK and with node- and path-based features (using a default linear kernel).

While the original study by \citeauthor{PMR1}~\cite{PMR1,PMR3} showed encouraging results, the design of PMR features was limited to features extracted from the methods' CFGs only. First generating CFGs from the source code and then extracting features is relatively costly as compared to extracting features directly from the source code. 
To see whether and how the PMR approach could be improved by using features extracted directly from source code, as well as using those features with other classification approaches than SVM, we decided to conduct a study using the same set of methods and the same set of MRs as in \citeauthor{PMR3} \cite{PMR3}. We complement the original PMR approach by looking at 21 features directly extracted from the source code. In addition, we experiment with five different binary classification models (including SVM). 

In the context of our study, we answer the following research questions:

\begin{itemize}
    \item \textbf{RQ$_{1}$:} What set of source code based features provides the best PMR performance?
    \item \textbf{RQ$_{2}$:} Does PMR performance improve when using source code based features instead of CFG-based features?
\end{itemize}

In our study, we focus on PMR feature engineering. In particular, we are interested in understanding whether and how it is possible to achieve a similar or better performance than that obtained by \citeauthor{PMR1}~\cite{PMR1, PMR3},  but using features extracted from source code rather than features extracted from the CFG.

The rest of the paper is structured as follows. \Cref{sec:RelWork} presents the related work. In \Cref{sec:methodology}, we describe the methodology. In \Cref{sec:results}, we present results and answers to our research questions. We discuss some threats to validity in \Cref{sec:discussion}. Finally, we conclude the paper in \Cref{sec:conclusion}.

\section{Related Work}
\label{sec:RelWork}

Since MT was introduced in 1998 by \citeauthor{chen2020metamorphic}~\cite{chen2020metamorphic}, it has been widely studied with increasing interest in recent years \cite{segura2016survey}. Several studies have shown MT as a strong technique for testing the ``non-testable programs" where an oracle is unavailable or too difficult to implement \cite{segura2016survey,10.1145/3143561,Murphy2008PropertiesOM,8573811,Murphy2008PropertiesOM}. Also, MT has been demonstrated to be an effective technique for testing in a variety of application domains, \textit{e.g.,} autonomous driving \cite{zhang2018deeproad,zhou2019metamorphic}, cloud and networking systems \cite{canizares2020mt,9477667}, bioinformatic software \cite{10.1145/3193977.3193981, shahri2019metamorphic}, scientific software \cite{peng2021contextual,8533366}. However, the efficacy of MT heavily relies on the specific MRs employed.

\citeauthor{PMR1}~\cite{PMR1}, were the first to show that, for previously unseen methods, applicable MRs can be predicted using ML techniques. Their work showed that classification models created using a set of features extracted from CFGs and a set of predefined MRs are effective in predicting whether a method in a newly developed SUT can be tested using a specific MR taken from the pre-defined set. Then, they extend their first work~\cite{PMR1} by conducting a feature analysis to identify the most effective CFG's related features for predicting MRs \cite{PMR3}. Their results showed that SVM models built with features based on CFG similarity measurements, in particular using RWK, perform better than SVM models using nodes- and paths-based features with linear kernel. 

\citeauthor{PMR4}~\cite{PMR4} extended the initial PMR study \cite{PMR1} using semi-supervised learning techniques on a set of node-based features and the CFG path tagged with six predefined MRs. \citeauthor{PMR5}~\cite{PMR5} applied PMR approach for predicting three high-level categories of MRs (\textit{i.e.,} Permutative, Additive, and Multiplicative) for matrix-based programs. Their results show that the RWK can effectively predict these MRs. \citeauthor{10.1145/3340482.3342741}~\cite{10.1145/3340482.3342741} explored and compared equivalent and non-equivalent mutants as data augmentation technique to broaden the training set using PMR. Their augmentation approach was tested on the PMR original study dataset~\cite{PMR1}. The study demonstrated that equivalent mutants are a valid data augmentation technique to improve the PMR detection rate. \citeauthor{8055540}~\cite{8055540} presented RBF-MLMR, a multi-label technique that predicts MRs using radial basis function neural networks. Instead of using several binary classifiers like in PMR, RBF-MLMR use a neural network to predict all potential MRs for a given method. The major difference between this technique and PMR is the usage of multi-label and neural networks, but it follows the same pipeline as PMR original study. Also, the RBF-MLMR's feature design is CFG's node- and path-based.

\section{Methodology}
\label{sec:methodology}

\begin{figure}[tp!]
	\centering
	\includegraphics[width=0.48\textwidth]{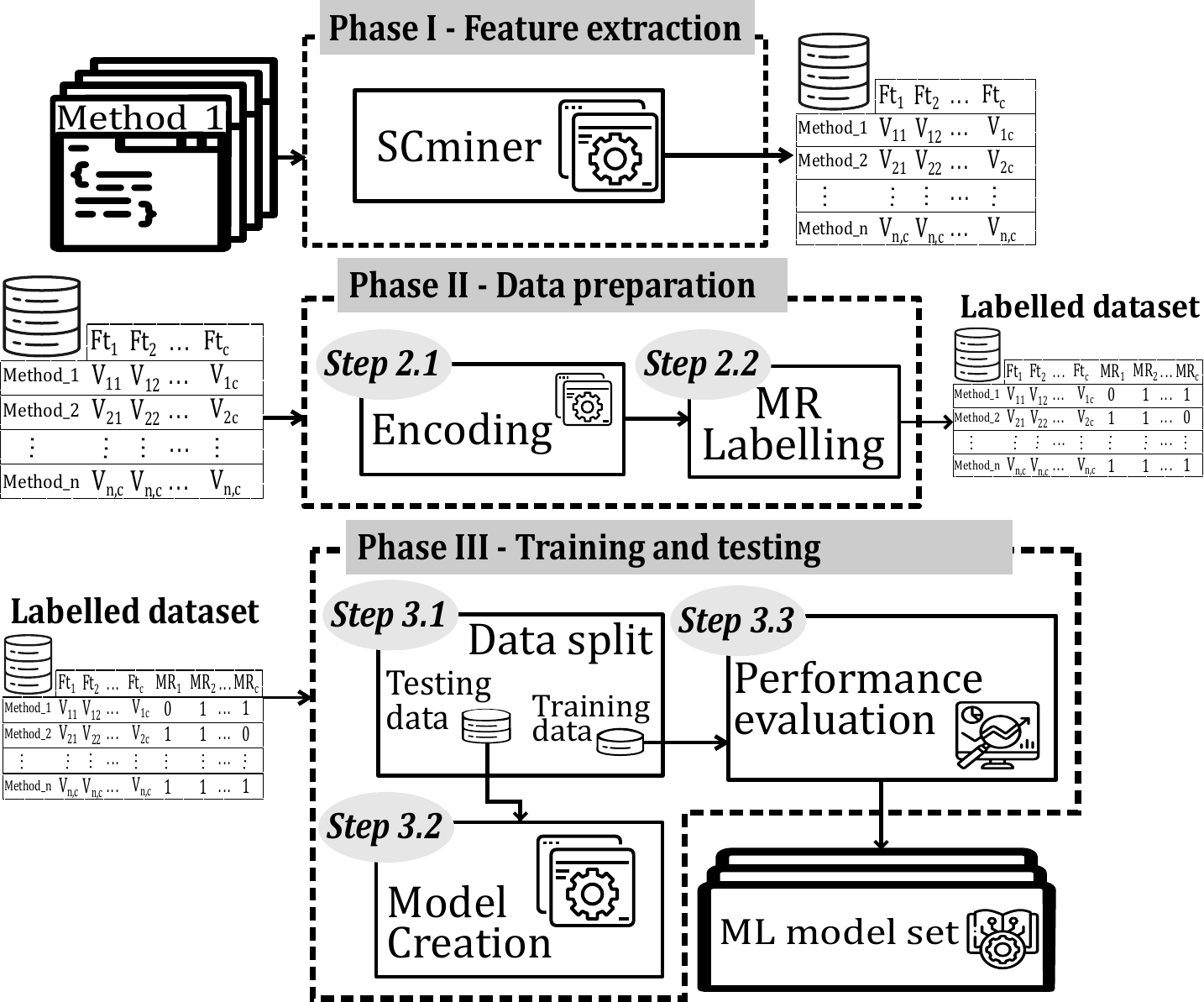}
	\caption{PMR procedure}
	\label{fig:ML-Model}
\end{figure}

In this paper, we focus on automated MR identification following the PMR approach but using the method's source code metrics instead of its CFG information. Answers to the research questions are obtained from the analysis of the results of several implementations of the PMR procedure. Therefore, in this section, we first present the PMR procedure, which is slightly different from the one proposed by \citeauthor{PMR1} \cite{PMR1, PMR3} (\Cref{subsec:PMR}). Then, we present the labelled dataset used in our study  (\Cref{subsec:dataset}). Finally, we present the performance measures used in this paper study (\Cref{subsec:PerformanceMeasures})

\subsection{PMR Procedure}
\label{subsec:PMR}
\Cref{fig:ML-Model} shows the PMR procedure. The PMR procedure consists of three phases. \textit{Phase I} is responsible for extracting metrics of the method source code. The output of this phase is a csv file with 21 source code metrics related features per method. \textit{Phase II} is in charge of preparing the data according to the requirements of the ML algorithms. For example, encoding categorical features. Also, each method is labelled with elements from the set of pre-defined MRs. \textit{Phase III} is in charge of training and evaluating the binary classification models that predict whether a specific MR is applicable to the unit testing of a specific method. Below we describe each phase in detail. 

\subsubsection{\textit{Phase I - Feature extraction}} in this phase, a list of features from various source code metrics is retrieved for each method. A source code metric is a quantitative measure of a software system's attribute. Measuring software complexity \cite{5385114}, assessing software maintainability \cite{303623}, measuring software quality \cite{10.5555/559784}, and other applications rely on source code metrics. Recent studies have demonstrated the usefulness of employing source code metrics in a variety of research areas, including defect prediction \cite{gao}, and time cost reduction in mutation testing \cite{9155985, 8730151, 8304576}. In our implementation, a total of 21 source code metrics are used, as shown in \Cref{tbl:softwareMetrics}. We use SCminer\footnote{https://github.com/aduquet/SCminer}, which is an open-source tool for mining source code metrics at method level that support three different programming languages (Java, C$++$, and Python).

\begin{table}[!h]
	\caption{Software metrics}
	\label{tbl:softwareMetrics}
	\centering
	\begin{tabular}{l|l}
		\toprule
		\textbf{Metric} & \textbf{Description}\\
		\midrule
		tloc & Total number of lines of code \\
		sloc\_whbl & Total number of code lines without blank lines \\
		\multirow{2}{*}{nloc} & Total number of code lines without comments or \\
		&libraries statements \\
		\multirow{2}{*}{nloc\_whbl} & Total number of code lines without blank lines, \\
		& comments or libraries statements \\ 
		sloc\_statements & Total number of lines with code statements \\
		token\_count &  Token is the word and operators, etc.\\
		start\_line & First code line\\
		end\_line & Last code line\\
		\multirow{2}{*}{full\_parameters} & It provides the full input parameters \\ 
		& including the variable name\\
		numArg & Number of inputs arguments\\
		dataArg & Inputs data type,  \textit{e.g., int array, int, float, etc.}\\
		numOper & Number of arithmetical operators\\
		\multirow{2}{*}{numOperands} & Number of operands, \textit{e.g., variables,}  \\
		&\textit{numeric and string constants} \\
		total\_Var & Total number of variables declared\\
		numLoops & Number of loops \\
		\multirow{2}{*}{CCN} & Cyclomatic Complexity Number, which is the number \\
		& of possible alternative paths through a piece of code \\
		numMethCall & Number of external methods called \\
		has\_return & Tells if the method has a return value\\
		totalReturn & Tells how many return statements the method has\\
		returnDataType & Return data type,  \textit{e.g., int array, int, float, etc.} \\
		ext & extention file (Java, C$++$, or Python)\\
		
		\bottomrule
	\end{tabular}
\end{table}

\subsubsection{Phase II -- Data preparation}
This phase is made up of two steps:

\textit{\textbf{Step 2.1 -- Encoding:} }This step is in charge of preparing the data according to the requirements of the ML algorithms. For instance, encoding categorical features. Also, a further feature selection analysis can be done. Feature selection is an extra activity that allows exploring the different performances of ML models when different features are used. 

\textit{\textbf{Step 2.2 -- Labelling MRs:} }Since PMR employs supervised learning classification techniques, which need the usage of a labelled dataset to give instances for learning. After the feature extraction phase and encoding step, the training dataset is constructed by manually labelling each method with suitable MRs. Depending on whether a certain MR does or does not satisfy the method, the method is labelled with a 1 or a 0 for such MR.

\subsubsection{\textbf{Phase III -- Training and testing}}
\label{subsubsec:PhaseIII}
This step entails extracting information from data using one or more supervised ML algorithms, or a combination of them. There are three steps that must be taken.

\textit{\textbf{Step 3.1 -- Data split:}} This step is responsible for splitting the dataset into two subsets: a training set and a test set. The training set is used to create the prediction model, while the test set is used to evaluate the performance of the created prediction model. 

\textbf{\textit{Step 3.2 -- Model creation}} refers to the process of building prediction models. Choosing a good modelling technique is vital for the training and prediction stage in any ML application, including the PMR approach. In this paper, a total of 5 popular classification algorithms are investigated when applying them to PMR, including:

\begin{itemize}
    \item Random Forest (RF)
    \item Support Vector Machine (SVM) with linear kernel
    \item Decision Trees (DT)
    \item Gaussian Naıve Bayes (GNB)
    \item Logistic Regression (LR)
\end{itemize}

\textit{\textbf{Step 3.3 -- Performance evaluation:}} This step measures the performance of the created prediction models. To assess classification performance, we utilised stratified $k$--fold cross\-validation. The $k$--fold cross\-validation approach assesses how well a prediction model performs on previously unknown data. The data set is randomly partitioned into $k$ subgroups in $k$--fold cross-validation. Then, $k$--1 subsets are utilised to develop the predictive model (training), and the remaining subset is used to assess the predictive model's performance (testing). This procedure is performed $k$ times, with each of the $k$ subsets being used to assess performance. In stratified $k$--fold cross-validation, $k$--folds are partitioned so that they include roughly the same percentage of positive (functions that display a specific MR) and negative (functions that do not exhibit a specific metamorphic relation) samples as the original data set.

\subsection{Dataset and pre-defined set of MRs}
\label{subsec:dataset}

\citeauthor{PMR3}~\cite{PMR3} provide a dataset with 100 Java methods in their CFG representation\footnote{http://www.cs.colostate.edu/saxs/MRpred/functions}. Instead of using the CFGs, we built a code corpus containing the source code of the same 100 Java methods. The methods are from the open-source libraries \textit{Colt Project}~\cite{colt}, an open-source library written for high-performance scientific and technical computing, \textit{Apache Mahou}~\cite{Mahout}, a machine learning library, \textit{Apache Commons Mathematics}~\cite{commons-maths}, a library of mathematics and statistics components, and \textit{Java Collections}~\cite{coll}, a framework that provides an architecture to store and manipulate the group of objects. All of these libraries are written in Java. To obtain the source code of the methods presented in the form of CFGs by \citeauthor{PMR3}, we search in the different libraries for the name of the method. 

To be able to make a fair comparison, we use the same set of six pre-defined MRs as in the original study \cite{PMR3}. \Cref{tbl:MR_specifications} lists the MRs used, the changes in the inputs and the expected outputs, and the total number of methods to which a specific MR applies. 
Details of the set of methods,\textit{ i.e.,} method name, library to which it belongs, and the MRs that apply, have been made available in github\footnote{\href{https://anonymous.4open.science/r/RENE-PredictingMetamorphicRelations-81C3/Methods/Methods.PNG}{Link to methods description}}.

\begingroup
\setlength{\tabcolsep}{6pt} 
\renewcommand{\arraystretch}{1} 
\begin{table}[ht!]
\centering
\caption{MRs used and total number of methods that match (\cmark) a specific MR}
{
	\label{tbl:MR_specifications}
	\resizebox{\linewidth}{!} {
	\begin{tabular}{l|l|l|l}
		\toprule
		\textbf{MR} & \textbf{Change in the input} & \textbf{Output expected} & \cmark   \\
		\toprule
	    ADD & Add a positive constant         & Increase or remain constant & $56$    \\
	    EXC & Remove an element               & Decrease or remain constant & $32$   \\
	    INC & Add a new element               & Increase or remain constant & $34$   \\
		MUL & Multiply by a positive constant & Increase or remain constant & $66$   \\
		PER & Permute the components          & Remain constant             & $33$   \\
		INV & Take the inverse of each element& Decrease or remain constant & $63$   \\
		\bottomrule
		\multicolumn{4}{l}{ADD: Additive, MUL: Multiplicative, PER: Permutative  } \\
		\multicolumn{4}{l}{INC: Inclusive, EXC: Exclusive, INV: Invertive  }
	\end{tabular}}}
\end{table}
\endgroup

\subsection{Performance Measures}
\label{subsec:PerformanceMeasures}
We evaluate the performance of our PMR implementation using prediction \textit{recall}, \textit{accuracy}, \textit{precision}, \textit{F1-score}, and \textit{AUC-ROC} for each subject using 10-fold cross validation and then summarise (using the mean) those statistics to compare the performance of different classification algorithms.

In this paper, we denote a classification output in which a specific $MR_{n}$ satisfies the method $m$ as the \textit{positive class} and a classification output in which specific $MR_{n}$ does not satisfy the method $m$ as the \textit{negative class}. Using this notation, each standard performance measure is expressed as a function of the counts of elements in the \emph{Confusion Matrix} defined as follows  (denote TP as true positive, TN as true negative, FP as false positive, FN as false negative): 

\begin{itemize}
    \item recall = TP / (TP + FN)
    \item accuracy =  (TP + TN) / (TP + FP + TN + FN)
    \item precision = TP / (TP + FP)
    \item F1-score is the harmonic mean of precision and recall
    \item AUC-ROC, area under the receiver operating characteristic (ROC) curve
\end{itemize}

\section{Results}
\label{sec:results}
The full set of data generated during our experiments as well as all scripts can be found in our GitHub repo\footnote{\href{https://anonymous.4open.science/r/VST22\_PMR-SourceCodeMetrics-E536/}{VST22\_PMR-SourceCodeMetrics-E536/}}.

\begin{table*}[ht!]
\caption{Feature importance score absolute values}
\label{tbl:featureImportance}
\centering
\resizebox{18cm}{!}{
\begin{tabular}{l|l|l|l|l|l|l|l|l|l|l|l|l|l|l|l|l|l|l|l|l|l }
    \toprule
    \textbf{MR}	&	
    \textbf{A} & \textbf{B} & \textbf{C} & \textbf{D} & \textbf{E} & \textbf{F} & \textbf{G} &
    \textbf{H} & \textbf{I} & \textbf{J} & \textbf{k} & \textbf{L} & \textbf{M} & \textbf{N} &
    \textbf{O} & \textbf{P} & \textbf{Q} & \textbf{R} & \textbf{S} & \textbf{T} & \textbf{U} \\
    \midrule
    
    ADD	&   0.92 & 1.00	&	0.71	&	0.40	&	0.38	&	0.31	&	0.30	&	0.30	&	0.29	&	0.29	&	0.28	&	0.28	&	0.27	&	0.26	&	0.24	&	0.23	&	0.22	&	0.21	&	0.00	&	0.00	&	0.00	\\
    EXC	&	1.00 & 0.86	&	0.62	&	0.54	&	0.48	&	0.48	&	0.46	&	0.45	&	0.45	&	0.43	&	0.42	&	0.30	&	0.28	&	0.26	&	0.24	&	0.24	&	0.22	&	0.21	&	0.00	&	0.00	&	0.00	\\
    INC	&	1.00 & 0.81	&	0.53	&	0.48	&	0.46	&	0.45	&	0.44	&	0.43	&	0.43	&	0.42	&	0.41	&	0.33	&	0.28	&	0.25	&	0.22	&	0.21	&	0.20	&	0.16	&	0.00	&	0.00	&	0.00	\\
    MUL	&	1.00 & 0.72	&	0.67	&	0.41	&	0.33	&	0.32	&	0.30	&	0.28	&	0.27	&	0.26	&	0.26	&	0.25	&	0.25	&	0.24	&	0.23	&	0.23	&	0.21	&	0.15	&	0.00	&	0.00	&	0.00	\\
    PER	&	0.57 & 1.00	&	0.54	&	0.41	&	0.38	&	0.36	&	0.24	&	0.23	&	0.23	&	0.22	&	0.22	&	0.21	&	0.20	&	0.19	&	0.18	&	0.15	&	0.14	&	0.09    &	0.00	&	0.00	&	0.00	\\
    INV	&	1.00 & 0.68	&	0.56	&	0.41	&	0.35	&	0.32	&	0.31	&	0.30	&	0.29	&	0.28	&	0.27	&	0.26	&	0.25	&	0.22	&	0.21	&	0.20	&	0.19	&	0.18	&	0.00	&	0.00	&	0.00	\\
	\midrule
	\midrule
	\textbf{AVG} &
	0.91 & 0.85	&	0.60	&	0.44	&	0.40	&	0.37	&	0.34	&	0.33	&	0.33	&	0.32	&	0.31	&	0.27	&	0.25	&	0.24	&	0.22	&	0.21	&	0.20	&	0.17	&	0.00	&	0.00	&	0.00	\\
	\bottomrule
	\multicolumn{22}{l}{\textbf{AVG:} Average, \textbf{A:} dataArg, \textbf{B:} CCN, \textbf{C:} tloc, \textbf{D:} sloc-whbl, \textbf{E:} sloc-statements, \textbf{F:} nloc-whbl, \textbf{G:} nloc, \textbf{H: }token-count, \textbf{I: }start-line, \textbf{}J: end-line, \textbf{K: }numArg} \\
	\multicolumn{22}{l}{ \textbf{L: n}umLoops,\textbf{ M: }totalVar,\textbf{	N: }numOper,	\textbf{O: }numMethCall,	\textbf{P: }hasReturn,
	\textbf{Q: }totalReturn,\textbf{	R: }numOperands,	\textbf{S: }returnDataType,	\textbf{T:} ext,	\textbf{U:} full-Parameters }
	\end{tabular}}
\end{table*}

\subsection{ RQ$_{1}$: What set of source code based features provides the best PMR performance?}
\label{subsec:RQ1_results}

\begin{figure}[ht!]
	\centering
	\includegraphics[width=0.49\textwidth, trim=3mm 3mm 9mm 1mm, clip ]{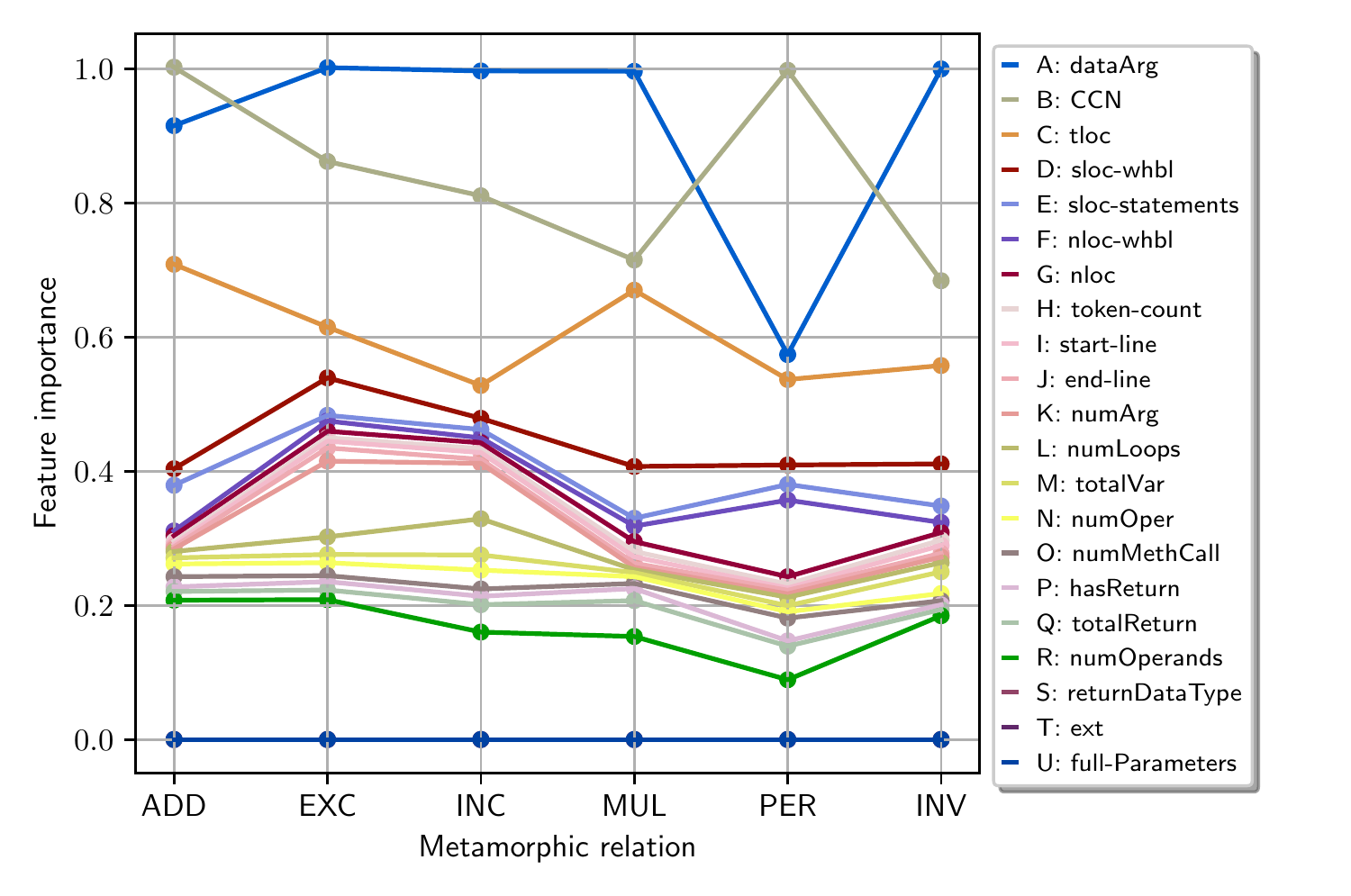}
	\caption{Feature importance score per MR}
	\label{fig:featureImportance}
\end{figure}

As a first step to answering RQ$_{1}$, we explore whether using a subset of the maximum set of 21 features helps improve the performance of each MR classifier. To do this, an RF classifier is used to calculate the importance score of each feature. Note that, taking randomness into account, we built RF classifiers ten times and then averaged the importance score for each feature. \Cref{fig:featureImportance} shows the feature importance score per MR and \Cref{tbl:featureImportance} lists the absolute importance values per MR. In both, \Cref{fig:featureImportance} and \Cref{tbl:featureImportance}, the features are ranked from highest to lowest score. 

From \Cref{fig:featureImportance} and \Cref{tbl:featureImportance} one can see that regardless of the MR, the importance ranking of features follows a fairly uniform pattern. It seems that from the third-ranked feature, \textit{i.e.,} “C: tloc",  regardless of the MR, features are ranked in the same order. Moreover, for the last three features, \textit{i.e., }“S: returnData Type", “T: ext", and “U: full-Parameters", the score is zero for all MRs. Also, it is important to highlight that from the fourth feature for the MRs: MUL, PER, and INV, their scores are quite close to each other for the same features. This happens also for the MRs: ADD, EXC, and INC, but from the eleventh feature, \textit{i.e.}, “L: numLoop", onward. Regarding the top-3 features with the best scores, the MRs: EXC, INC, MUL and INV follow the same pattern concerning the order of the feature importance score, this is “A: dataArg", “B: CCN" and “C: tloc". In MRs: ADD and PER, the order of importance differ because “B: CCN" is the highest score instead of “A: dataArg". It could be said that this is the only exception to the general ranking pattern we found.

\begingroup
\setlength{\tabcolsep}{6pt} 
\renewcommand{\arraystretch}{1} 
\begin{table}[ht!]
\centering
\caption{AUC-ROC and precision absolute values per MR using the top-ranked $n$ features in a RF classifier }
{
	\label{tbl:featureImportanceAUCROC}
	\resizebox{\linewidth}{!} {
	\begin{tabular}{l|l|l|l|l|l|l|l|l}
		\toprule
		\multirow{2}{*}{\textbf{Metric}}& \multirow{2}{*}{\textbf{MR}} & \multicolumn{7}{c}{\textbf{Top-ranked $n$ features}}  \\
		& & Feat$_{3}$ & Feat$_{6}$ & Feat$_{9}$ & Feat$_{12}$ & Feat$_{15}$ & Feat$_{18}$ & Feat$_{21}$ \\
		\toprule
        \multirow{6}{*}{\textbf{AUC$^\star$}}  
        & ADD	&	0.620	&	0.590	&	0.620	&	\textbf{0.886}	&	0.620	&	0.590	&	0.590	\\
        & EXC	&	0.629	&	0.593	&	0.583	&	\textbf{0.833}	&	0.589	&	0.623	&	0.666	\\
        & INC	&	\textbf{0.720}	&	0.675	&	\textbf{0.720}	&	0.717	&	0.675	&	0.675	&	0.694	\\
        & MUL	&	0.649	&	0.518	&	0.518	&	\textbf{0.762}	&	0.518	&	0.491	&	0.578	\\
        & PER	&	\textbf{0.763}	&	0.725	&	0.725	&	0.747	&	0.725	&	0.641	&	0.747	\\
        & INV	&	0.595	&	0.611	&	0.588	&	0.625	&	0.636	&	0.545	&	\textbf{0.640}	\\
        \midrule
        \multirow{6}{*}{\textbf{Prec$^\pm$}}  
        & ADD	&	0.627	&	0.613	&	0.640	&	0.767	        &	0.740	&	0.729	&	\textbf{0.769}	\\
        & EXC	&	0.667	&	0.651	&	0.693	&	\textbf{0.866}	&	0.667	&	0.688	&	0.614	\\
        & INC	&	0.767	&	0.733	&	0.733	&	\textbf{0.888}	&	0.727	&	0.761	&	0.652	\\
        & MUL	&	\textbf{0.875} & 0.833	&	0.854 &	0.853	&	0.675	&	0.630	&	0.657	\\
        & PER	&	0.625	&	0.761	&	0.805	&	\textbf{0.814}	&	0.625	&	0.625	&	0.625	\\
        & INV	&	0.675	&	0.625	&	0.714	&	\textbf{0.833}	&	0.75	&	0.600	&	0.675	\\

		\bottomrule
		\multicolumn{7}{l}{\textbf{Feat:} Feature, $^\star$\textbf{AUC:} AUC-ROC, $^\pm$\textbf{Prec:} Precision } \\
	\end{tabular}}}
\end{table}
\endgroup

After ranking the features from highest to lowest according to their importance score, we explored the gradual selection of best-ranked features. Also, we attempted to discover the best subset in the reduced subject set using 10-fold cross-validation and RF classifier. We started with a subset with the three best-ranked features; then we augment the subset with the following three features, and so on. In total, we evaluated seven subsets for each MR. Each time, the next three highest-ranked features were added until reaching the total set of 21 features.

\Cref{tbl:featureImportanceAUCROC} shows the performance values obtained using the RF classifier in terms of AUC-ROC and Precision when the classifier is trained with a different feature subsets. In this analysis, we want to know which subset of features is the best based on the AUC-ROC and Precision performance measures. AUC-ROC indicates the extent to which the model can distinguish between classes. The higher the AUC-ROC, the better the model will predict the positive class as positive and the negative class as negative. In our context, the positive case is the one when an MR applies to a method. Since we plan to use the PMR approach for generating initial test cases for methods that are yet lacking tests, it is more important that we avoid the occurrence false positives. A false positive would result in generating tests based on a MR that actually is not applicable. The performance measure \textit{Precision} is commonly used to assess the capability of a binary classifier to predict the positive case correctly. Therefore, we are particularly interested in this measure when comparing the performance of the various classifiers we build.

With regards to AUC-ROC, \Cref{tbl:featureImportanceAUCROC} shows that three out of six MRs, i.e., ADD, EXC, and MUL, have the best results with the top-ranked 12 features, with a difference greater than 0.2 to all other feature sets. For MRs INC and PER, the highest values for AUC-ROC are achieved when the 3 top-ranked features are used. However, the difference to the performance using 12 top-ranked features is less than about 0.01. For MR INV, the best AUC-ROC score is when the complete set of features is used. However, as for INC and PER, the difference with the 12 top-ranked features is rather small (0.015). Therefore, based on the AUC-ROC measure, it seems to be reasonable to simply use the 12 top-ranked features across the board for all MRs.

With regards to \textit{Precision}, \Cref{tbl:featureImportanceAUCROC} shows that for four out of six MRs, i.e., EXC, INC, PER and INV, the best performance is achieved with the 12 top-ranked features. The highest precision value for the MR ADD is obtained when all 21 features are used. However, the difference between 21 features and 12 top-ranked features is only 0.002, approximately. This indicates that performance-based on precision does not vary significantly if the top 12 features are used.

In the next step, we explore the PMR approach using four different classification models in addition to RF, i.e., DT, GNB, SVM and LG. Each classifier is trained with the top-ranked 3 and 12 features, and the full set of 21 features. We include the sets of 3 and 21 features in the analysis to double-check whether the choice of 12-features is also the best for other classifiers than RF. We evaluate the performance of each classifier using 10-fold cross-validation. Specifically, 70\% of the instances’ datasets are used for training in each fold, and the remaining 30\% are used for testing. The prediction statistics (\textit{accuracy, precision, recall, F1 score, and AUC-ROC}) are recorded for each fold. The average values of these statistics for each classifier and each MR are listed in \Cref{tbl:RQ1_final}.

\begin{table*}[!tp]
\caption{PMR performance metrics when using 3, 12, and 21 top-ranked features on RF, DT, GNB, SVM, and LG classifiers}
\label{tbl:RQ1_final}
\centering
\resizebox{\textwidth}{!}{
\begin{tabular}{l|l|lll|lll|lll|lll|lll}
    \toprule
    \multirow{2}{*}{\textbf{MR}}
    &\multirow{2}{*}{\textbf{Clas$^\star$}}
    &\multicolumn{3}{c|}{\textbf{Accuracy}} & \multicolumn{3}{c|}{\textbf{Precision}} & \multicolumn{3}{c|}{\textbf{Recall}} & \multicolumn{3}{c|}{\textbf{F1 score}} & \multicolumn{3}{c}{\textbf{AUC-ROC}} \\
    
    & 
    & Feat$_{3}$ & Feat$_{12}$ & Feat$_{21}$ 
    & Feat$_{3}$ & Feat$_{12}$ & Feat$_{21}$
    & Feat$_{3}$ & Feat$_{12}$ & Feat$_{21}$
    & Feat$_{3}$ & Feat$_{12}$ & Feat$_{21}$
    & Feat$_{3}$ & Feat$_{12}$ & Feat$_{21}$\\
 
    \midrule
    \multirow{5}{*}{\textbf{ADD}}  
    & RF	
    &	0.764	&	\textbf{0.886}	&	0.884	
    &	0.627	&	0.767	&	\textbf{0.769}	
    &	0.972	&	0.971	&	\textbf{0.973}	
    &	0.827	&	\textbf{0.830}	&	0.824	
    &	0.620	&	\textbf{0.886}	&	0.590	\\
    
    & DT	
    &	0.787	&	\textbf{0.800}	&	0.680	
    &	0.769	&	\textbf{0.812}	&	0.725	
    &	0.750	&	\textbf{0.833}	&	0.666	
    &	0.692	&	\textbf{0.727}	&	0.656	
    &	0.757	&	\textbf{0.816}	&	0.698	\\
    
    & GNB	
    &	0.809	&	\textbf{0.833}	&	0.726	
    &	0.805	&	\textbf{0.914}	&	0.695	
    &	0.780	&	\textbf{0.833}	&	0.726	
    &	0.796	&	\textbf{0.823}	&	0.769	
    &	0.771	&	\textbf{0.875}	&	0.667	\\
    
    & SVM	
    &	0.740	&	\textbf{0.803}	&	0.620	
    &	0.754	&	\textbf{0.833}	&	0.675	
    &	0.729	&	\textbf{0.838}	&	0.620	
    &	0.747	&	\textbf{0.835}	&	0.659	
    &	0.753	&	\textbf{0.833}	&	0.673	\\
    
    & LG	
    &	0.680	&	\textbf{0.800}	&	0.680	
    &	0.750	&	\textbf{0.807}	&	0.693	
    &	0.762	&	\textbf{0.833}	&	0.690	
    &	0.762	&	\textbf{0.833}	&	0.690	
    &	0.738	&	\textbf{0.800}	&	0.675	\\

    \midrule
    \multirow{5}{*}{\textbf{EXC}}  
    & RF	
    &	0.725	&	\textbf{0.800}	&	0.650	
    &	0.667	&	\textbf{0.866}	&	0.614	
    &	0.833	&	\textbf{1.000}	&	0.666	
    &	0.733	&	\textbf{0.800}	&	0.666	
    &	0.629	&	\textbf{0.833}	&	0.666	\\
    
    & DT
    &	0.660	&	\textbf{0.750}	&	0.570	
    &	0.770	&	\textbf{0.844}	&	0.695	
    &	0.600	&	\textbf{0.667}	&	0.532	
    &	0.551	&	\textbf{0.602}	&	0.500	
    &	0.602	&	\textbf{0.667}	&	0.536	\\
    
    & GNB	
    &	0.690	&	\textbf{0.712}	&	0.667	
    &	0.733	&	\textbf{0.800}	&	0.666	
    &	0.600	&	\textbf{0.667}	&	0.534	
    &	0.564	&	\textbf{0.667}	&	0.461	
    &	0.595	&	\textbf{0.619}	&	0.571	\\
    
    & SVM	
    &	0.672	&	\textbf{0.704}	&	0.640
    &	0.700	&   \textbf{0.733}	&	0.667	
    &	0.599	&	\textbf{0.667}	&	0.530	
    &	0.710	&	\textbf{0.857}	&	0.562	
    &	0.634	&	\textbf{0.694}	&	0.573	\\
    
    & LG	
    &	0.662	&	0.657	&	\textbf{0.667}	
    &	0.735	&	\textbf{0.844}	&	0.625	
    &	0.628	&	\textbf{0.665}	&	0.590	
    &	0.621	&	\textbf{0.671}	&	0.571	
    &	0.610	&	\textbf{0.696}	&	0.523	\\

	\midrule
    \multirow{5}{*}{\textbf{INC}} 
	&	RF	
	&	0.765	&	\textbf{0.900}	&	0.630	
	&	0.767	&	\textbf{0.888}	&	0.652	
	&	0.778	&	\textbf{0.890}	&	0.666	
	&	0.808	&	\textbf{0.888}	&	0.727	
	&	\textbf{0.720}	&	0.717	&	0.694	\\
	
    &	DT	
    &	0.681	&	\textbf{0.750}	&	0.612	
    &	0.566	&	\textbf{0.616}	&	0.516	
    &	0.465	&	\textbf{0.596}	&	0.333	
    &	0.513	&	\textbf{0.589}	&	0.438	
    &	0.564	&	\textbf{0.605}	&	0.523	\\
    
    &	GNB	
    &	0.687	&   \textbf{0.705}	&	0.668	
    &	0.733	&	\textbf{0.862}	&	0.604	
    &	0.627	&	\textbf{0.667}	&	0.587	
    &	0.604	&	\textbf{0.667}	&	0.542	
    &	0.596	&	\textbf{0.681}	&	0.510	\\
    
    &	SVM	
    &	0.706	&	\textbf{0.802}	&	0.610	
    &	0.737	&	\textbf{0.806}	&	0.667	
    &	0.590	&	\textbf{0.645}	&	0.534	
    &	0.558	&	\textbf{0.667}	&	0.448	
    &	0.655	&	\textbf{0.761}	&	0.548	\\
    
    &	LG	
    &	0.657	&	\textbf{0.701}	&	0.613	
    &	0.648	&	\textbf{0.695}	&	0.601	
    &	0.633	&	\textbf{0.668}	&	0.597	
    &	0.452	&	\textbf{0.571}	&	0.333	
    &	0.560	&	\textbf{0.690}	&	0.429	\\

	\midrule
    \multirow{5}{*}{\textbf{MUL}}
    &	RF	
    &	0.725	&	\textbf{0.800}	&	0.650	
    &	\textbf{0.875}	&	0.853	&	0.657	
    &	0.912	&	\textbf{1.000}	&	0.823	
    &	0.812	&	\textbf{0.875}	&	0.748	
    &	0.649	&	\textbf{0.762}	&	0.578	\\
    
    &	DT	
    &	0.662	&	\textbf{0.703}	&	0.620	
    &	0.749	&	\textbf{0.802}	&	0.695	
    &	0.828	&	\textbf{0.833}	&	0.822	
    &	0.776	&	\textbf{0.833}	&	0.719	
    &	0.721	&	\textbf{0.791}	&	0.651	\\  
    
    &	GNB	
    &	0.680	&	\textbf{0.700}	&	0.660	
    &	0.755	&	\textbf{0.834}	&	0.675	
    &	0.899	&	\textbf{0.940}	&	0.857	
    &	0.803	&	\textbf{0.823}	&	0.782	
    &	0.575	&	\textbf{0.625}	&	0.524	\\
    
    &	SVM	
    &	0.707	&	\textbf{0.804}	&	0.610	
    &	0.792	&	\textbf{0.850}	&	0.733	
    &	0.845	&	\textbf{0.857}	&	0.833	
    &	0.742	&	\textbf{0.857}	&	0.626	
    &	0.726	&	\textbf{0.761}	&	0.690	\\
    
    &	LG	
    &	0.741	&	\textbf{0.801}	&	0.680	
    &	0.772	&	\textbf{0.875}	&	0.669	
    &	0.840	&	\textbf{0.857}	&	0.823	
    &	0.817	&	\textbf{0.875}	&	0.759	
    &	0.792	&	\textbf{0.833}	&	0.750	\\

	\midrule
    \multirow{5}{*}{\textbf{PER}} 
    
    &	RF	
    &	0.715	&	\textbf{0.810}	&	0.620	
    &	0.625	&	\textbf{0.814}	&	0.675	
    &	0.639	&	\textbf{0.712}	&	0.566	
    &	0.726	&	\textbf{0.789}	&	0.662	
    &	\textbf{0.763}	&	0.725	&	0.747	\\
    
    &	DT	
    &	0.708	&	\textbf{0.750}	&	0.666	
    &	0.846	&	\textbf{0.914}	&	0.777	
    &	0.721	&	\textbf{0.775}	&	0.666	
    &	0.717	&	\textbf{0.857}	&	0.576	
    &	0.809	&	\textbf{0.857}	&	0.761	\\
    
    &	GNB	
    &	0.623	&	\textbf{0.700}	&	0.545	
    &	0.725	&	\textbf{0.828}	&	0.622	
    &	0.500	&	\textbf{0.666}	&	0.333	
    &	0.589	&	\textbf{0.727}	&	0.450	
    &	0.604	&	\textbf{0.642}	&	0.566	\\
    
    &	SVM	
    &	0.875	&	\textbf{0.910}	&	0.840	
    &	0.837	&	\textbf{0.875}	&	0.799	
    &	0.709	&	\textbf{0.750}	&	0.667	
    &	0.698	&	\textbf{0.729}	&	0.667	
    &	0.793	&	\textbf{0.825}	&	0.761	\\
    
    &	LG	
    &	0.765	&	\textbf{0.830}	&	0.700	
    &	0.803	&	\textbf{0.822}	&	0.783	
    &	0.688	&	\textbf{0.709}	&	0.667	
    &	0.720	&	\textbf{0.750}	&	0.690	
    &	0.745	&	\textbf{0.795}	&	0.694	\\

	\midrule
    \multirow{5}{*}{\textbf{INV}} 
	&	RF	
	&	0.655	&	\textbf{0.702}	&	0.608	
	&	0.675	&	\textbf{0.833}	&	0.675	
	&	0.788	&	\textbf{0.857}	&	0.719	
	&	0.776	&	\textbf{0.800}	&	0.751	
	&	0.595	&	0.625	&	\textbf{0.640}	\\
	
    &	DT	
    &	0.762	&	\textbf{0.800}	&	0.600	
    &	0.703	&	\textbf{0.844}	&	0.563	
    &	0.762	&	\textbf{0.857}	&	0.667	
    &	0.691	&	\textbf{0.714}	&	0.667	
    &	0.604	&	\textbf{0.667}	&	0.541	\\
    
    &	GNB	
    &	0.661	&	\textbf{0.701}	&	0.620	
    &	0.659	&	\textbf{0.833}	&	0.484	
    &	0.759	&	\textbf{0.857}	&	0.660	
    &	0.787	&	\textbf{0.823}	&	0.750	
    &	0.568	&	\textbf{0.625}	&	0.511	\\
    
    &	SVM	
    &	0.776	&	\textbf{0.802}	&	0.750	
    &	0.768	&	\textbf{0.844}	&	0.692	
    &	0.799	&	\textbf{0.833}	&	0.764	
    &	0.813	&	\textbf{0.857}	&	0.769	
    &	0.762	&	\textbf{0.857}	&	0.667	\\
    
    &	LG	
    &	0.768	&	\textbf{0.860}	&	0.676	
    &	0.728	&	\textbf{0.761}	&	0.695	
    &	0.794	&	\textbf{0.857}	&	0.731	
    &	0.767	&	\textbf{0.857}	&	0.676	
    &	0.714	&	\textbf{0.762}	&	0.667	\\

	\bottomrule
	\multicolumn{10}{l}{$^\star$Classifier}
	\end{tabular}}
\end{table*}

As expected, \Cref{tbl:RQ1_final} confirms that using the 12 top-ranked features is almost always the best choice across the board (best performance is printed in bold). In those cases where 12 features don't yield the best performance, the difference to the best performing feature set is alway less than 0.02. 

To decide which classifier is the best for each MR, we relied on the values of AUC-ROC and \textit{Precision} when using classifiers based on 12 features. 

For MR ADD, the highest average of AUC-ROC and \textit{Precision} is achieved when using GNB (average of 0.875 and 0.914). 

For MR EXC, the highest average of AUC-ROC and \textit{Precision} is achieved when using RF (average of 0.833 and 0.866). 

For MR INC, the highest average of AUC-ROC and \textit{Precision} is achieved when using RF (average of 0.717 and 0.888).

For MR MUL, the highest average of AUC-ROC and \textit{Precision} is achieved when using LG (average of 0.833 and 0.875).

For MR PER, the highest average of AUC-ROC and \textit{Precision} is achieved when using DT (average of 0.857 and 0.914).

For MR INV, the highest average of AUC-ROC and \textit{Precision} is achieved when using SVM with linear kernel (average of 0.857 and 0.844).

In summary, the results show that there is not one best classifier. RF is best for two MRs and each of the other classifiers is best for exactly one MR.


\begin{tcolorbox}
With regards to RQ$_1$, our results indicate that we achieve almost always a performance greater than 0.8 in terms of AUC-ROC and \textit{Precision} when predicting MRs using source code based features (the one exception is the case of AUC-ROC for MR INC). The best results are obtained with only 12 features out of 21. However, there is not one single best classifier for all MRs. With regards to \textit{Precision}, each classifier is best for one MR at least once. With regards to AUC-ROC, with the exception of DT, each classifier is best for one MR at least once.
\end{tcolorbox}

\subsection{RQ$_{2}$: Does PMR performance improve when using source code based features instead of CFG-based features?}
\label{subsec:RQ2_results}

The left-hand side of \Cref{tbl:RQ2} shows the AUC-ROC values obtained when SVM models are trained with three feature extraction approaches, i.e., node and path features (NF-PF), graphlet kernel (GK), and random walk kernel (RWK), as reported by \citeauthor{PMR3}~\cite{PMR3}. The features used in these classifiers are CFG-related. Since precision has not been reported by \citeauthor{PMR3}, we must base our comparison exclusively on the AUC-ROC measure. Also, \citeauthor{PMR3}\ do not report results from other models but SVM. 
The right-hand side of \Cref{tbl:RQ2} shows the AUC-ROC values for the five classifiers using 12 top-ranked source code based features. The highest AUC-ROC values for each MR are printed in bold font.

\begingroup
\setlength{\tabcolsep}{6pt} 
\renewcommand{\arraystretch}{1} 
\begin{table}[ht!]
\centering
\caption{Comparison between AUC-ROC values per MR obtained by \citeauthor{PMR3}~\cite{PMR3} using SVM with CFG-related features (NF-PF, GK and RWK), and AUC-ROC values obtained when using RF, DT, GNB and LG, with the 12 top-ranked source code based features. }
{
	\label{tbl:RQ2}
	\resizebox{\linewidth}{!} {
	\begin{tabular}{l|l|l|l|l|l|l|l|l}
		\toprule
		\multirow{3}{*}{\textbf{MR}}& \multicolumn{3}{c|}{\textbf{\citeauthor{PMR3}~\cite{PMR3}}}& \multicolumn{5}{c}{\multirow{2}{*}{\textbf{12 top-ranked source code based feat.}}}  \\
		&  NF-PF & GK & RWK  \\
		& SVM & SVM & SVM & SVM & RF & DT & GNB & LG\\
		\toprule
        ADD	&	0.81	&	0.83	&	\textbf{0.92}	&	0.83	&	0.89	&	0.82	&	0.88	&	0.80	\\
        EXC	&	0.78	&	0.78	&	\textbf{0.90}	&	0.69	&	0.83	&	0.67	&	0.62	&	0.70	\\
        INC	&	0.84	&	0.88	&	\textbf{0.89}	&	0.76	&	0.72	&	0.61	&	0.68	&	0.69	\\
        MUL	&	0.73	&	0.78	&	\textbf{0.83}	&	0.76	&	0.76	&	0.79	&	0.63	&	\textbf{0.83}	\\
        PER	&	0.93	&	0.91	&	\textbf{0.95}	&	0.83	&	0.76	&	0.86	&	0.64	&	0.80	\\
        INV	&	0.84	&	0.68	&	0.76	&	\textbf{0.86}	&	0.64	&	0.67	&	0.63	&	0.76	\\

		\bottomrule
		\multicolumn{7}{l}{\textbf{Feat:} Feature, $^\star$\textbf{AUC:} AUC-ROC, $^\pm$\textbf{Prec:} Precision } \\
	\end{tabular}}}
\end{table}
\endgroup

\begin{figure}[ht!]
	\centering
	\includegraphics[width=0.49\textwidth, trim=3mm 3mm 6mm 1mm, clip ]{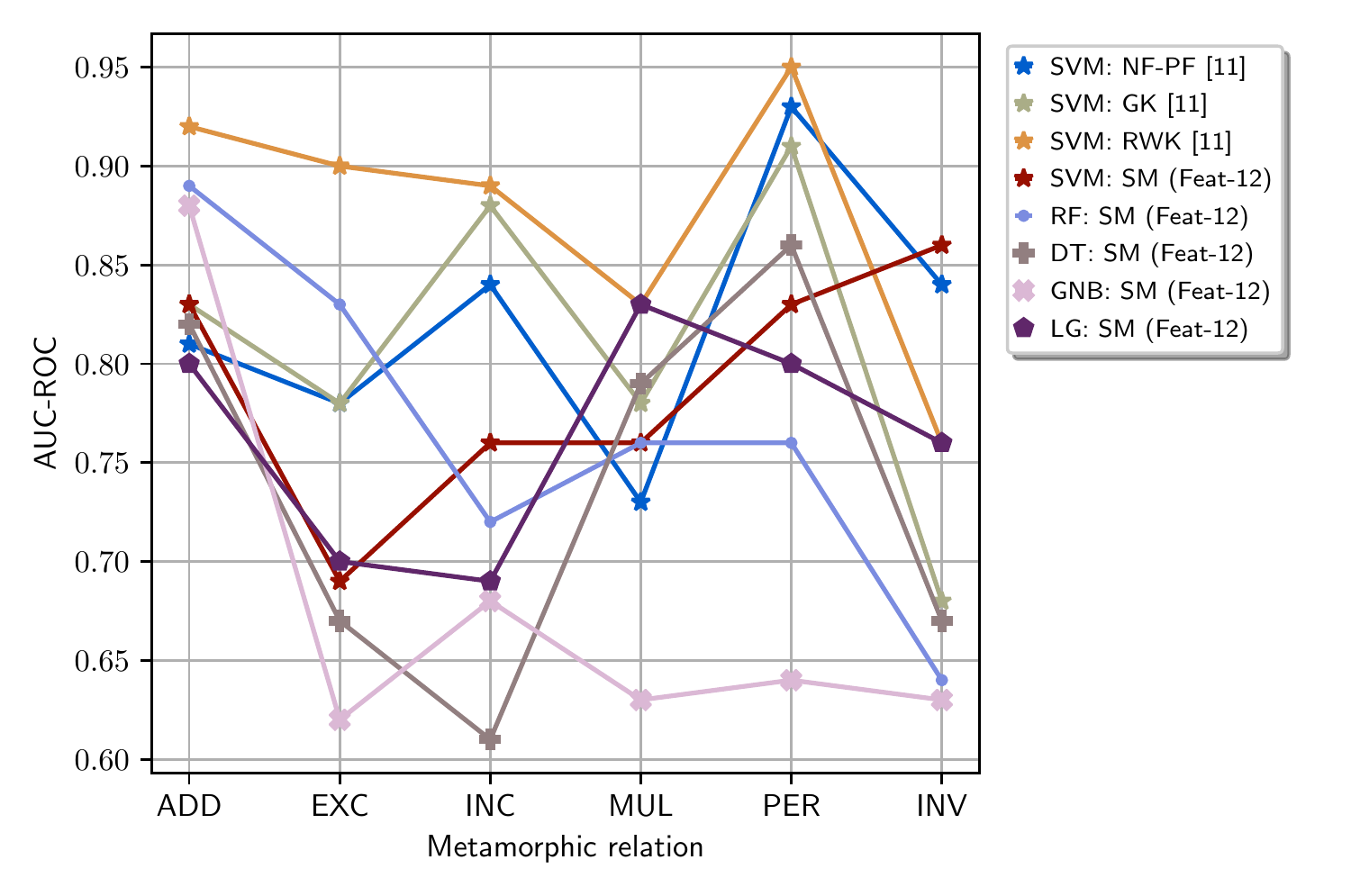}
	\caption{Comparison of the AUC-ROC results obtained by \citeauthor{PMR3}~\cite{PMR3} using node- path-based features (NF-PF), Graphlet Kernel (GK) and Random Walk Kernel (RWK) in SVMs, and the AUC-ROC results obtained in this work using the source code Metrics (SM) in SVM, RF, DT, GNB, and LG}
	\label{fig:NF_PFvsGKvsRWKvsSF}
\end{figure}

The comparison between \citeauthor{PMR3} and our models shows that in five out of six cases \citeauthor{PMR3}'s RWK-SVM model is performing best. Only for MR INV our SVM model using 12 top-ranked source code based features and linear kernel is performing best. For MR MUL, our LG classifier achieves a tie. However, considering that the extraction of features from CFGs is more expensive than building classifiers directly from the source code (i.e., without first having to construct CFGs from the source code and then analyse them), our classifiers might still be acceptable if their performance is not much lower than that of \citeauthor{PMR3}'s best classifier. From \Cref{fig:NF_PFvsGKvsRWKvsSF} (which plots the values presented in \Cref{tbl:RQ2}), one can see that not only for MR INV one of our classifiers performs best but also for all other MRs one of our classifiers has a performance close to that of \citeauthor{PMR3}'s best classifier. 





\begin{tcolorbox}
Regarding RQ$_2$, our results indicate that classifiers using source code based features most of the time cannot achieve better performance in terms of AUC-ROC. SVM classifiers using CFG-based features and the RWK are best for five out of six MRs. Only for INV, the SVM classifier using twelve source code based features (and the default linear kernel) is better than the SVM classifier with RWK proposed by \citeauthor{PMR3}~\cite{PMR3}. However, the performance of classifiers using source code based features is not dramatically worse than that of the best classifier using CFG-based features. Since source code based features are much cheaper to extract, this might outweigh the small loss of performance.
\end{tcolorbox}

\section{Discussion}
\label{sec:discussion}
We now discuss the four most relevant threats to validity of our study.

\subsection{Internal validity}
\label{subsec:Internal_validity}
Since \citeauthor{PMR3}'s original study only reported the performance of their classifiers only in terms of AUC-ROC, we had to base our comparison on that measure although, for our intended application of the PMR approach, precision would be the more appropriate measure. In addition, more studies are needed to investigate how it affects the overall performance of PMR when the same method is implemented differently, as this would directly affect feature extraction in both the original study and our modelling approach.

\subsection{External validity}
\label{subsec:External_validity}
For the sake of fair comparison, our study uses the same set of methods as the original study but uses the source code of the method instead of its CFG representation. However, it would have been preferable to use a corpus of source code consisting of a greater number of methods. Consequently, both our study and the original one cannot determine the true extent of the efficacy of the PMR approach.

\subsection{Construct validity}
\label{subsec:construc_validity}
In this paper, we used the SCmine framework to extract the source code metrics (features) at method level. SCminer is an open-source tool that uses third-party libraries. Usage of these third-party libraries represents potential threats to construct validity. To avoid this, we verified that the results produced by SCminer are correct by manually inspecting randomly selected outputs produced by the tool. 

\subsection{Conclusion validity}
\label{subsec:conclusionValidity}
We used AUC-ROC and presicion value for evaluating the performance of the classifiers. We considered AUC-ROC and Presicion $>$ 0.80 as a good classifier. This is consistent with most of the ML literature.

\section{Conclusion}
\label{sec:conclusion}
In this paper, we evaluate the performance of PMR using features related to the source code. We start by extracting 21 metrics related to the source code as features. Next, we perform features importance analysis using RF classifiers. After selecting the best set of features, we evaluate them in five different classifiers to find out which is the best in terms of AUC-ROC and Precision. Finally, to see if source code-related features improve PMR performance when using CFG-related features; we compared the AUC-ROC results obtained by \citeauthor{PMR3}~\cite{PMR3} with our own results. In summary, a total of 21 characteristics and 5 classification algorithms are evaluated in this study. All classifiers are carefully evaluated using 10-time cross-validation; To evaluate the performance of our PMT implementation, 5 performance metrics are recorded (per fold) and then averaged. Our results show that PMR can achieve results greater than 0.8 in terms of precision when predicting MRs using features based on the source code. The best results are obtained with only 12 features out of 21. However, there is no single best classifier for all MRs. Also, classifiers that use source-based features most of the time cannot perform better in terms of AUC-ROC. For this particular performance metric, the use of CFG-related functions in particular RWK is better than ours four out of six times and ties once.

\section*{Acknowledgement}
This research was partly funded by the Estonian Center of Excellence in ICT research (EXCITE), the European Regional Development Fund, the IT Academy Programme for ICT Research Development, the Austrian ministries BMVIT and BMDW, the Province of Upper Austria under the COMET (Competence Centers for Excellent Technologies) program managed by FFG, and grant PRG1226 of the Estonian Research Council.

\balance
\printbibliography

\end{document}